\documentclass[%
 aip,
 amsmath,amssymb,
 reprint,%
]{revtex4-1}

\usepackage{graphicx}% Include figure files
\usepackage{dcolumn}% Align table columns on decimal point
\usepackage{bm}% bold math
\usepackage[utf8]{inputenc}
\usepackage[T1]{fontenc}
\usepackage{mathptmx}
\usepackage{etoolbox}
\usepackage{xcolor}

\makeatletter
\def\@email#1#2{%
 \endgroup
 \patchcmd{\titleblock@produce}
  {\frontmatter@RRAPformat}
  {\frontmatter@RRAPformat{\produce@RRAP{*#1\href{mailto:#2}{#2}}}\frontmatter@RRAPformat}
  {}{}
}%
\makeatother
\begin{document}

\preprint{AIP/123-QED}

\title[Edge modes for flexural waves]{Edge modes for flexural waves in quasi-periodic linear arrays of scatterers }
% Force line breaks with \\
\author{Marc Mart\'i-Sabat\'e}
\author{Dani Torrent}%
\email{dtorrent@uji.es}

\affiliation{GROC, UJI, Institut de Noves Tecnologies de la Imatge (INIT), Universitat Jaume I, 12071, Castell\'o, (Spain)}

\date{\today}% It is always \today, today,
\begin{abstract}
We present a multiple scattering analysis of robust interface states for flexural waves in thin elastic plates. We show that finite clusters of linear arrays of scatterers built on a quasi-periodic arrangement support bounded modes in the two-dimensional space of the plate. The spectrum of these modes plotted against the modulation defining the quasi-periodicity has the shape of a Hofstadter butterfly, which previous works suggest that might support topologically protected modes. Some interface states appear inside the gaps of the butterfly, which are enhanced when one linear cluster is merged with its mirror reflected version. The robustness of these modes is verified by numerical experiments in which different degrees of disorder are introduced in the scatterers, showing that neither the frequency nor the shape of the modes is altered. Since the modes are at the interface between two one-dimensional arrays of scatterers deposited on a two-dimensional space, these modes are not fully surrounded by bulk gaped materials, so that they are more suitable for their excitation by propagating waves. The generality of these results goes beyond flexural waves, since similar results are expected for acoustic or electromagnetic waves.
\end{abstract}

\maketitle
%%%%%%%%%%%%%%%%%%%%%%%%%%%%%%%%%%%%%%%%%%%%%%%%%%%%%%%%%%%%%%%%%%%%%%%%
\section{\label{sec:intro}Introduction}
%%%%%%%%%%%%%%%%%%%%%%%%%%%%%%%%%%%%%%%%%%%%%%%%%%%%%%%%%%%%%%%%%%%%%%%%
The control and localization of mechanical waves is one of the most fundamental problems in phononics, since managing the energy carried out by these waves is important for a plethora of applications like cloaking, focusing, imaging or energy harvesting. The limitations of natural materials to achieve this control were overcome by the so called phononic crystals and metamaterials, conceived as artificially structured materials whose properties can be easily tailored\cite{ma2016acoustic,jin2019gradient}. 

More recently, with the advent of topological materials in condensed matter physics\cite{qi2008topological,schnyder2008classification}, new and exciting phases of matter have been discovered with remarkable properties. Among others, the existence of edge states robust against disorder is one of the most interesting from the point of view of wave propagation, therefore classical analogues of these states have received increasing attention\cite{haldane2008possible,prodan2009topological,wang2009observation,wu2015scheme,fleury2016floquet,ni2019observation}.

In acoustics and elasticity, topologically protected edge states have been studied in a wide variety of periodic and quasi-periodic materials\cite{yang2015topological,he2016acoustic,jin2018robustness,apigo2019observation,jin2020topological,rosa2021exploring,acin2018quantum,van2021bending,acin2018quantum,apigo2018topological,mitchell2018amorphous,ungureanu2021localizing,kuznetsova2021localized}. When the interface state occurs in a two or three dimensional space we have a one or two dimensional interface, respectively, where the field can propagate without suffering back-scattering, while if it happens in a one-dimensional space the interface state is a topologically protected zero-dimensional bound mode, although recently protected states have been found in two-dimensional domains by means of the classical analogue to the Majorana fermion\cite{chen2019mechanical,gao2019majorana,wen2022topological}. However, all these  states are surrounded by the bulk material, so that their excitation might require propagation outside the domain of interest or penetration through a gaped material. 

In this work we give a step forward towards the design of localized interface modes in mechanical systems. We have considered a quasi-periodic line of scatterers embedded in a two-dimensional elastic plate. We have applied multiple scattering theory to the study of these structures, which is a reliable tool for the analysis of finite structures\cite{packo2019inverse,marti2021dipolar,packo2021metaclusters} against common methods based on super-cells, since these introduce some artifacts due to periodicity that are not obvious in some occasions. We have shown that bound modes appear in the line of scatterers when these are rigid enough, and that the spectrum of these modes follows the well-known Hofstadter butterfly in the appropriate space. We have found that edge states appear in the gaps of the butterfly for finite clusters and that when a cluster is placed together with its mirror reflected version the existence of these states is enhanced, in the sense that their quality factor is higher. We have shown as well that these modes are robust against positional disordering of the scatterers, robustness verified with multiple scattering simulations. The advantage of these modes is that they are zero dimensional modes, trapped between two one-dimensional ``bulk'' materials in a two-dimensional space, which is a great advantage from the practical point of view, since the bound state is not fully surrounded by gaped bulk structures. 

%%%%%%%%%%%%%%%%%%%%%%%%%%%%%%%%%%%%%%%%%%%%%%%%%%%%%%%%%%%%%%%%%%%%%%%%
\section{\label{sec:mst}Bounded modes in linear clusters of scatterers}
%%%%%%%%%%%%%%%%%%%%%%%%%%%%%%%%%%%%%%%%%%%%%%%%%%%%%%%%%%%%%%%%%%%%%%%%

\begin{figure}[h!]
	\centering
	\includegraphics[width=\linewidth]{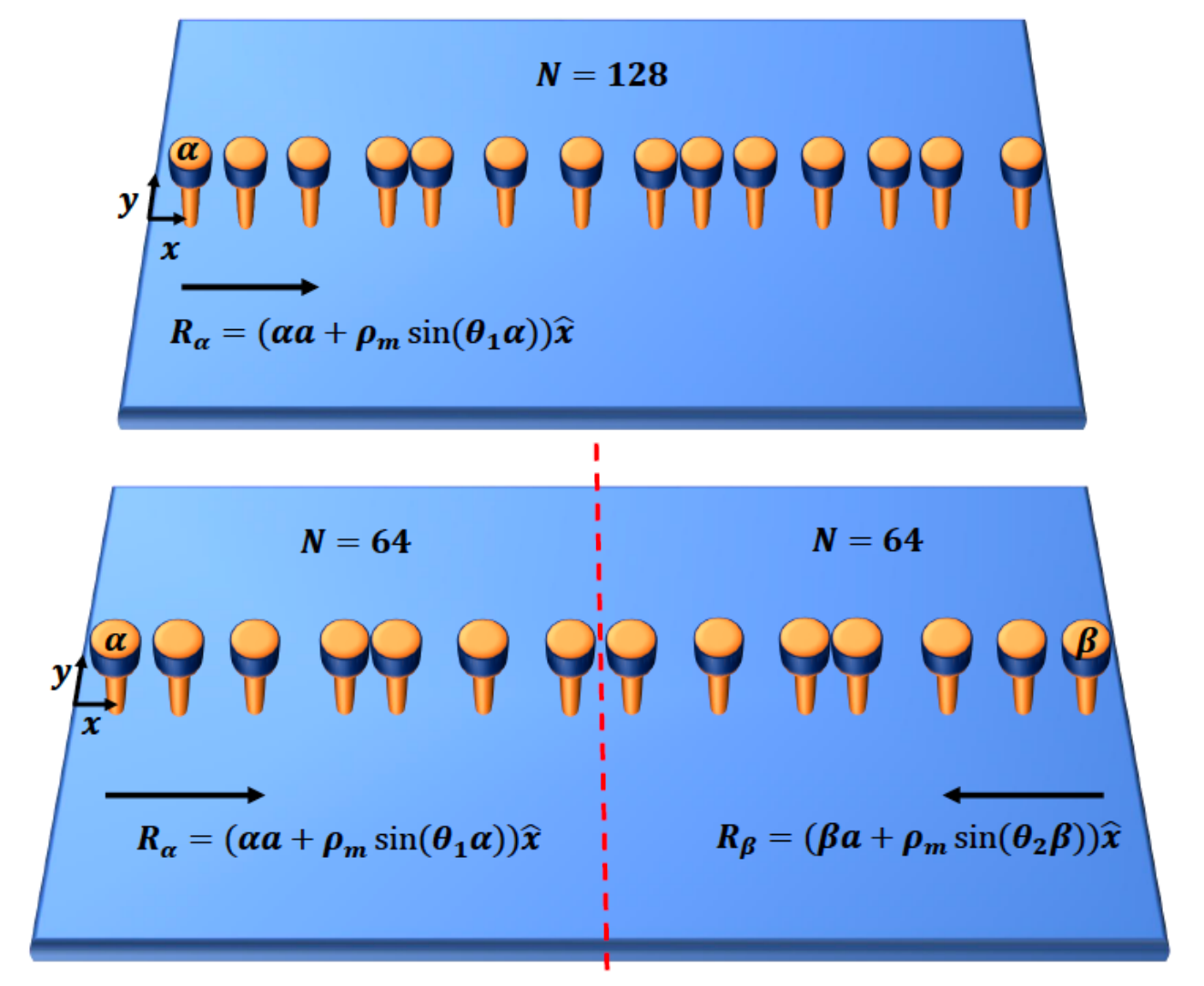}
	\caption{Schematic diagram of the two geometries explored in the text. Upper panel shows a quasi-periodic line of scatterers in a thin elastic plate, while lower panel shows a quasi-periodic line merged with its mirror symmetric version.}
	\label{Figure1}
\end{figure}

Let us assume that we have a cluster of $N$ point scatterers attached to a thin elastic plate (see figure \ref{Figure1}, upper panel, for a schematic view) in positions $\mathbf{R}_{\alpha}$, for $\alpha = 1,2,...,N$. In this work we will assume that these scatterers are arranged in a linear quasi-periodic distribution such that the position of the $\alpha$ scatterer is\cite{apigo2019observation}
\begin{equation}
    \mathbf{R}_{\alpha} = a\alpha + \rho_m\sin(\alpha\theta),
\end{equation}
where $a$ is the lattice constant, $\rho_m$ is the radius of the modulation circle and $\theta$ is the angle rotated in the circle (as defined in [\onlinecite{apigo2019observation}]). An infinite cluster ($N = \infty$) is periodic whenever $\theta/2\pi \in \mathbb{Q}$. The number of scatterers that form a period is $P$ and can be found applying the condition $P\theta/2\pi \in \mathbb{Z}$.

The propagation and scattering of time-harmonic flexural waves of frequency $\omega$, in the plate is described by its vertical displacement $W=\psi e^{-i\omega t}$ which satisfies a multiple scattering wave equation\cite{torrent2013elastic}
\begin{equation}
\label{eq:W1}
    (\nabla^4-\omega^2\rho h/D)\psi=\sum_\alpha t_\alpha \delta(\mathbf{r}-\mathbf{R}_\alpha)\psi,
\end{equation}
where $D$ bending stiffness, $\rho$ is the mass density and $h$ is the thickness of the plate. The characteristic impedance $t_{\alpha}$ of the scatterers is defined as
\begin{equation}
    t_\alpha=\gamma_\alpha \frac{\Omega^2\Omega_\alpha^2}{\Omega_\alpha^2-\Omega^2},
\end{equation}
with
\begin{equation}
\gamma_\alpha=\frac{m_\alpha}{\rho a^2 h},
\end{equation}
being $m_\alpha$ the mass of the scatterers and $\Omega$ and $\Omega_\alpha$ the operating frequency and their resonant frequency, respectively, in reduced units, which are defined as
\begin{equation}
    \Omega^2=\omega^2\frac{\rho a^2 h }{D}.
\end{equation}
In the above two equations $a$ is an arbitrary unit of length, defined for its suitability when studying periodic materials but, as can be easily seen, the impedance $t_\alpha$ is actually independent of $a$. If equation \eqref{eq:W1} is multiplied by the parameter $a^4$, we obtain 
\begin{equation}
\label{eq:W2}
    (a^4\nabla^4-\Omega^2a^2)\psi=\sum_\alpha a^4 t_\alpha \delta(\mathbf{r}-\mathbf{R}_\alpha)\psi,
\end{equation}
where it is easy to see that lengths are normalized with respect $a$ and frequency is given in units of $\Omega a$. In the right hand side,  the two-dimensional delta function absorbes $a^2$ and the remaining $a^2$ is introduced in $t_\alpha$, which now is given in terms of $\Omega a$ and $\Omega_\alpha a$,
\begin{equation}
    t_\alpha=\gamma_\alpha \frac{\Omega^2a^2\Omega_\alpha^2a^2}{\Omega_\alpha^2a^2-\Omega^2a^2},
\end{equation}
Consequently, to perform numerical experiments, the only required parameters are $a$, $\Omega a$ and the set of $t_\alpha$. These normalized units simplify the understanding of the underlying physics of the problem.

The solution to the multiple scattering problem when some external field $\psi_0(\mathbf{r})$ impinges on the cluster, consists of this incident field plus a scattered field, so that the total field is\cite{torrent2013elastic}
\begin{equation}
    \psi(\mathbf{r}) = \psi_0(\mathbf{r}) + \sum_{\alpha=1}^N B_{\alpha}G(\mathbf{r}-\mathbf{R}_{\alpha}). 
\end{equation}

The $B_\alpha$ coefficients are obtained self-consistently from the familiar multiple scattering system of equations
\begin{equation}
\label{eq:mst}
    \sum_{\beta = 1}^N M_{\alpha\beta}B_{\beta} = \psi_0(\mathbf{R}_{\alpha}),
\end{equation}
where the matrix elements $M_{\alpha\beta}$ are given by
\begin{equation}
    M_{\alpha\beta} = t_{\alpha}^{-1}\delta_{\alpha\beta} - G(\mathbf{R}_{\alpha} - \mathbf{R}_{\beta}),
\end{equation}
with $G(\mathbf{r})$ being the Green's function of the flexural wave equation,
\begin{equation}
    G(\mathbf{r})=\frac{i}{8k^2_b}\left[H_0(k_br)+\frac{2i}{\pi}K_0(k_br)\right],
\end{equation}
where $H_0(\cdot)$ is the zero-order Hankel function, $K_0(\cdot)$ is the zero-order modified Bessel function of the second kind and being $k_b$ the wavenumber of the incident field, related with frequency as
\begin{equation}
    k_b^4a^4=\Omega^2a^2,
\end{equation}
 
The eigenfrequencies of the cluster are found as the non-trivial solutions of the system of equations \eqref{eq:mst} when there is no incident field, which is equivalent to find those frequencies for which the determinant of the $M$ matrix is zero or to find an eigenvalue of the matrix equal to zero, which is a more suitable method from the numerical point of view. This happens only for complex frequencies if the cluster is finite, but a good approximation can be found by analyzing the minimum eigenvalue $\lambda_{\min}$ of $M$, as was previously done in reference [\onlinecite{marti2021dipolar}]. This parameter will never be zero for real frequencies, however it can be assumed that if a strongly localized mode appears in the cluster, the difference between an open and a closed system will be very small, which in turn means that a local minimum of $\lambda_{\min}$ is expected near the real part of the resonant frequency. The role of the imaginary part of the frequency will be to completely cancel this eigenvalue, thus we can asume that the smaller $\lambda_{\min}$ for a real frequency, the smaller the imaginary part of the resonance and, therefore, the better the quality of the mode.

\begin{figure}[h!]
	\centering
	\includegraphics[width=\linewidth]{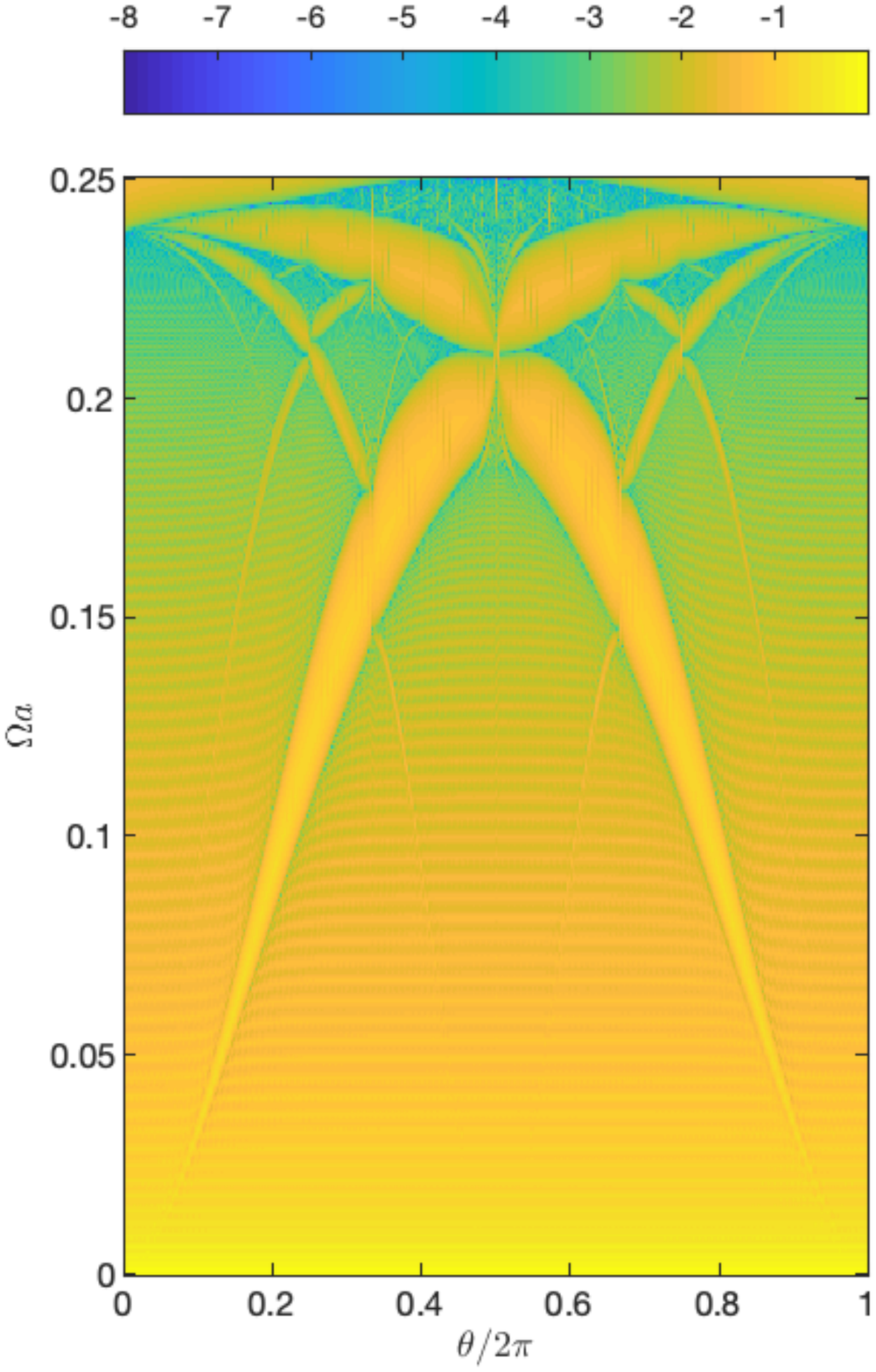}
	\caption{Evolution of the minimum eigenvalue of the multiple scattering matrix as a function of both the modulation parameter $\theta$ and the frequency. The fractal diagram that appears in the map is the well-known Hofstadter's butterfly.}
	\label{Figure2}
\end{figure}

Figure \ref{Figure2} shows a map of the minimum eigenvalue of $M$ for a cluster of $N=$ 128, assuming $a=1$, $\gamma_\alpha= 100$, $\Omega_{\alpha}a= 0.25$ and $\rho_m=0.5$. The plot shows function $f=f(\Omega,\theta)$ defined as 
\begin{equation}
\label{eq:log}
f=\log_{10}|\lambda_{\min}(\Omega,\theta)|
\end{equation}
so that high negative values (blue points) show the existence of an eigenmode. As we see, the diagram forms the well-known Hofstadter's butterfly. This structure is characterized by the opening of several gaps without modes all over the spectrum, defining the contour of the butterfly (yellow regions in the figure). As it was expected, the system is symmetric to $\theta/2\pi = 0.5$. Moreover, tuning the $\rho_m$ parameter these gaps can be broadened or narrowed. At the resolution shown in this figure it is not possible to distinguish the existence of modes inside the gaps, but a zoomed version of this plot shows them, as will be discussed in next section.

%%%%%%%%%%%%%%%%%%%%%%%%%%%%%%%%%%%%%%%%%%%%%%%%%%%%%%%%%%%%%%%%%%%%%%%%
\section{\label{sec:interface}Interface states in quasi-periodic clusters of scatterers}
%%%%%%%%%%%%%%%%%%%%%%%%%%%%%%%%%%%%%%%%%%%%%%%%%%%%%%%%%%%%%%%%%%%%%%%%
Interface states appear at the edges of finite structures which present band gaps in their spectrum. In the case of quasi-periodic structures, recent works suggest that the gaps shown in Hofstadter's butterfly could be topological\cite{apigo2019observation}, so that robust edge states are expected at the interface.

\begin{figure}[h!]
	\centering
	\includegraphics[width=\linewidth]{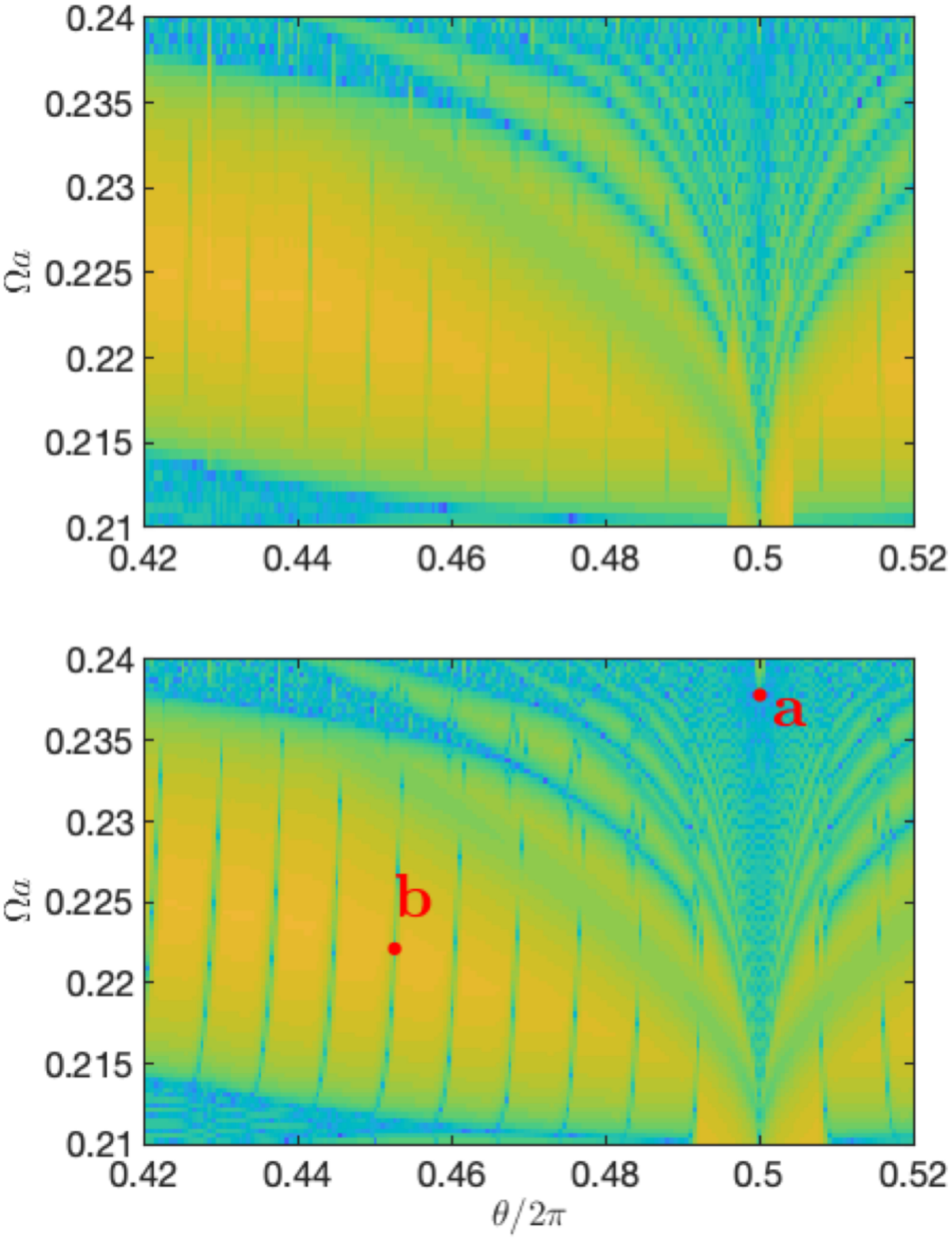}
	\caption{Upper panel: zoom of figure \ref{Figure2} showing interface states inside the gaps of the butterfly. Lowe panel: Same figure but for two faced clusters with different modulation parameter $\theta$. $x$ axis states the modulation parameter for the first cluster, while the second one can be obtained as $2\pi-\theta$. We see how the interface states are clearly more visible what is indicative of a better localization.}
	\label{Figure3}
\end{figure}

Figure \ref{Figure3}, upper panel, shows a zoomed region of figure \ref{Figure2}, where we can see some week blue modes inside the gaps of the butterfly, interpreted as edge states of the finite cluster. Lower panel shows equation \eqref{eq:log} but this time the structure is formed by two different modulated clusters, whose modulation pattern begins at the edge of the structure, forming an interface at the centre of the cluster. Both clusters are quasi-periodic with $N=64$, but the left cluster is built with a given $\theta$ parameter while the right cluster is its specular reflection (see figure \ref{Figure1}, lower panel). As we can see, the structure of the butterfly is identical to the upper panel but now the interface states have been enhanced, since the blue regions defining the modes are more defined and, as discussed before, this is indicative of a higher quality of the mode. The reason is that, in the single cluster configuration, the state is located at the edge of the cluster and it is surrounded by the plate's free space, so that it will be more leaky than in the second case where the interface is sandwiched between two linear clusters, what will improve the quality factor of the mode. 

\begin{figure}[h!]
	\centering
	\includegraphics[width=\linewidth]{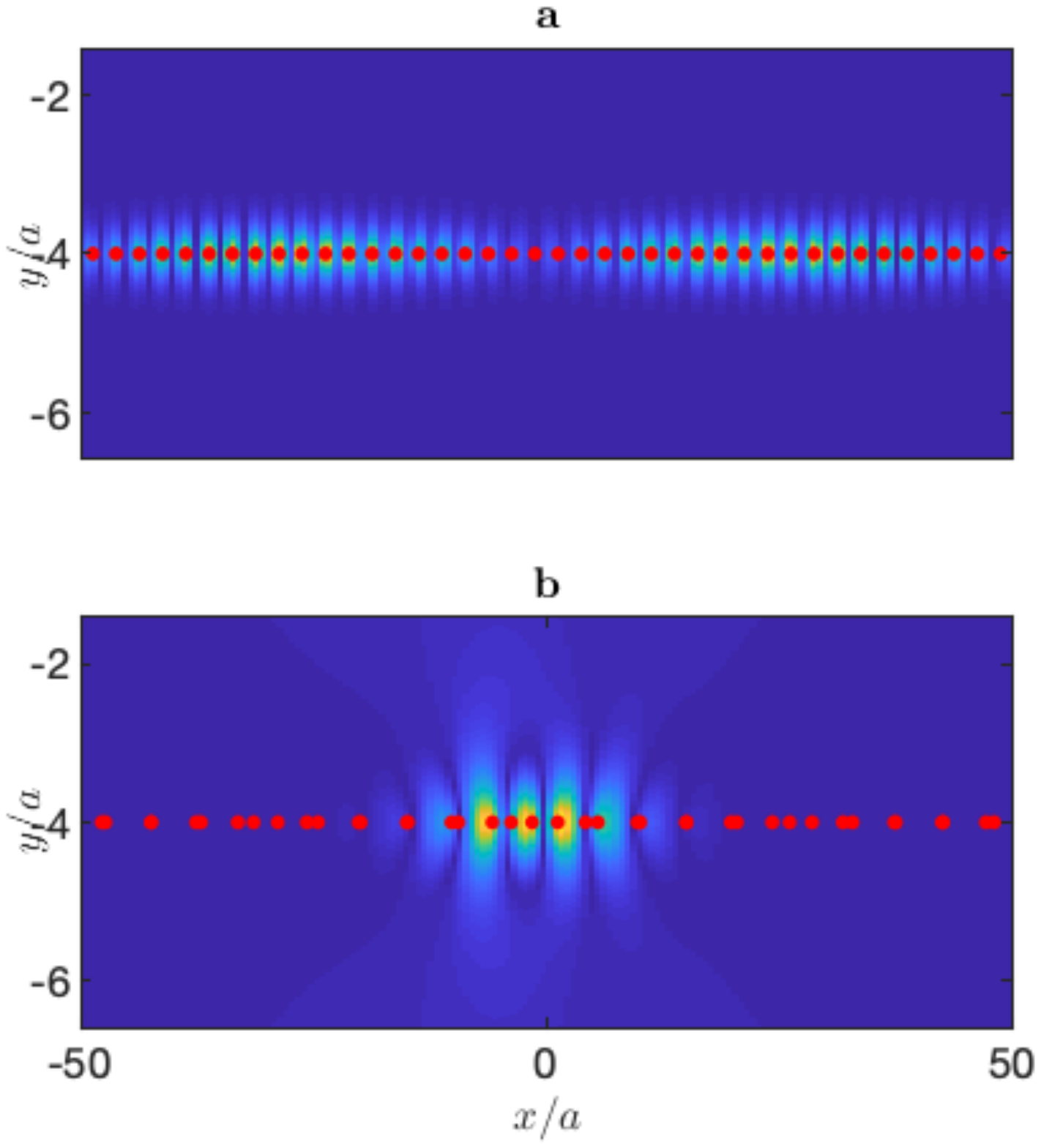}
	\caption{Spatial distribution of the pointed modes shown in Figure \ref{Figure3}. Only the central scatterers of the cluster are shown. $\mathbf{a}$ is the minimum eigenvalue for the whole range of frequencies of the periodic cluster with $\theta/2\pi = 0.5$. As it can be seen, this mode propagates all along the cluster. $\mathbf{b}$ is a cluster mode located inside of the bandgap. Its scattering field is localized at the center of the cluster, where the change of the modulation parameter if found.}
	\label{Figure4}
\end{figure}

An example of the different modes found in the previous analysis is depicted in figure \ref{Figure4} (modes labeled by ``a'' and ``b'' in the lower panel of figure \ref{Figure3}). Panel ``a'' shows a mode corresponding to the periodic configuration $\theta/2\pi=0.5$; we see how the field is not localized at any specific point but it is distributed all along the cluster. The hot spots correspond to the classical profile of a standing wave trapped in a finite waveguide, in this case the periodic array of scatterers. However, panel ``b'' corresponds to a configuration with $\theta/2\pi=0.4525$, the aperiodic cluster is merged with its specular reflection at $x=0$ and we see how a localized mode appears at the interface between the two clusters.

We see therefore that the periodic finite line of scatterers supports bound states, as it is indeed a closed waveguide and, therefore, defines a resonant cavity. However, introducing quasi-periodicity increases the number of modes, and the spectrum maps a Hofstadter's butterfly as a function of the modulation defining the quasi-periodicity. When these clusters are merged with their mirror version, the quality of the interface states is enhanced and they can be easily observed.

%%%%%%%%%%%%%%%%%%%%%%%%%%%%%%%%%%%%%%%%%%%%%%%%%%%%%%%%%%%%%%%%%%%%%%%%
\section{\label{sec:random}Robustness against positional disorder}
\begin{figure}[h!]
	\centering
	\includegraphics[width=\linewidth]{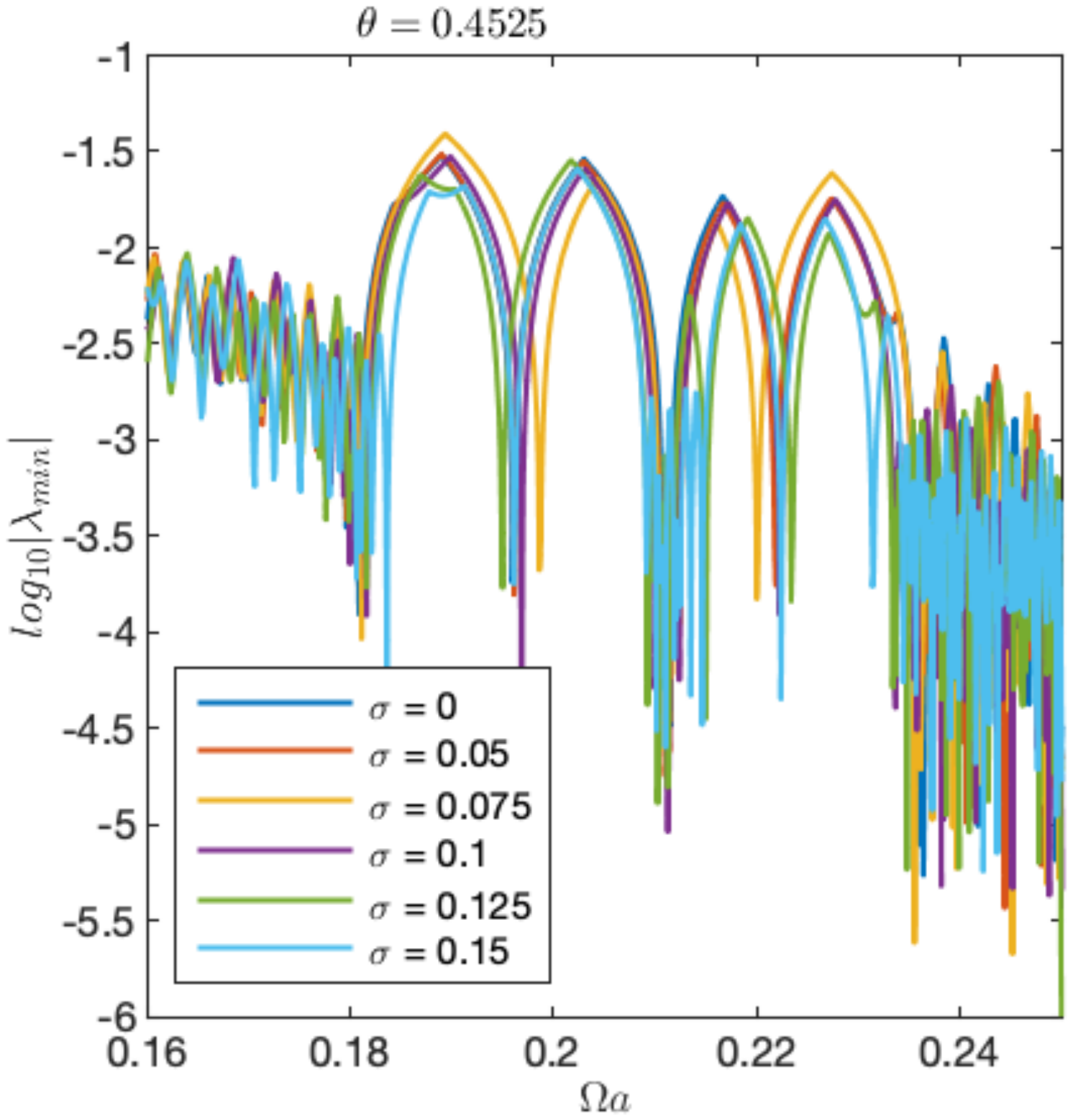}
	\caption{Minimum eigenvalue for a faced cluster with $\theta_1/2\pi = 0.4525$ and $\theta_2/2\pi = 0.5475$. Each line corresponds to a cluster with a different degree of disorder, characterized by $\sigma$. At both sides of $\Omega a = 0.21$ we have two localized modes. The effect of disorder in this system makes a shift on the localized mode frequency; however, the mode is still presence in the cluster. Thus, it can be stated that these edge states are robust.}
	\label{Figure5}
\end{figure}

%%%%%%%%%%%%%%%%%%%%%%%%%%%%%%%%%%%%%%%%%%%%%%%%%%%%%%%%%%%%%%%%%%%%%%%%
The most interesting feature of edge states is their topological protection, i.e., their robustness against small perturbations. In order to check the robustness of the edge modes found in the previous section we have performed several numerical experiments with multiple scattering theory. In our experiments we add ``positional disorder'' to the clusters, so that, for every scatterer $\alpha$ in the cluster, we perform a perturbation to its position such that now 
\begin{equation}
    \mathbf{R}_\alpha^\sigma=\mathbf{R}_\alpha+\sigma a Z
\end{equation}
with $Z$ being a normal random variable of unitary variance and zero mean. The parameter $\sigma$ characterizes the amount of disorder, since it ensures that all the scatterers are deviated from its initial position a quantity that is normally distributed between $-3\sigma a$ and $3\sigma a$. 
Figure \ref{Figure5} shows the eigenvalue function defined in equation \eqref{eq:log} for the two-clusters configuration for $\theta_1/2\pi=0.4525$  and for different amounts of disorder characterized by $\sigma$. We can see how the edge modes found are robust, since they remain only slightly shifted in frequency when we increase the disorder parameter $\sigma$. 

\begin{figure}[h!]
	\centering
	\includegraphics[width=\linewidth]{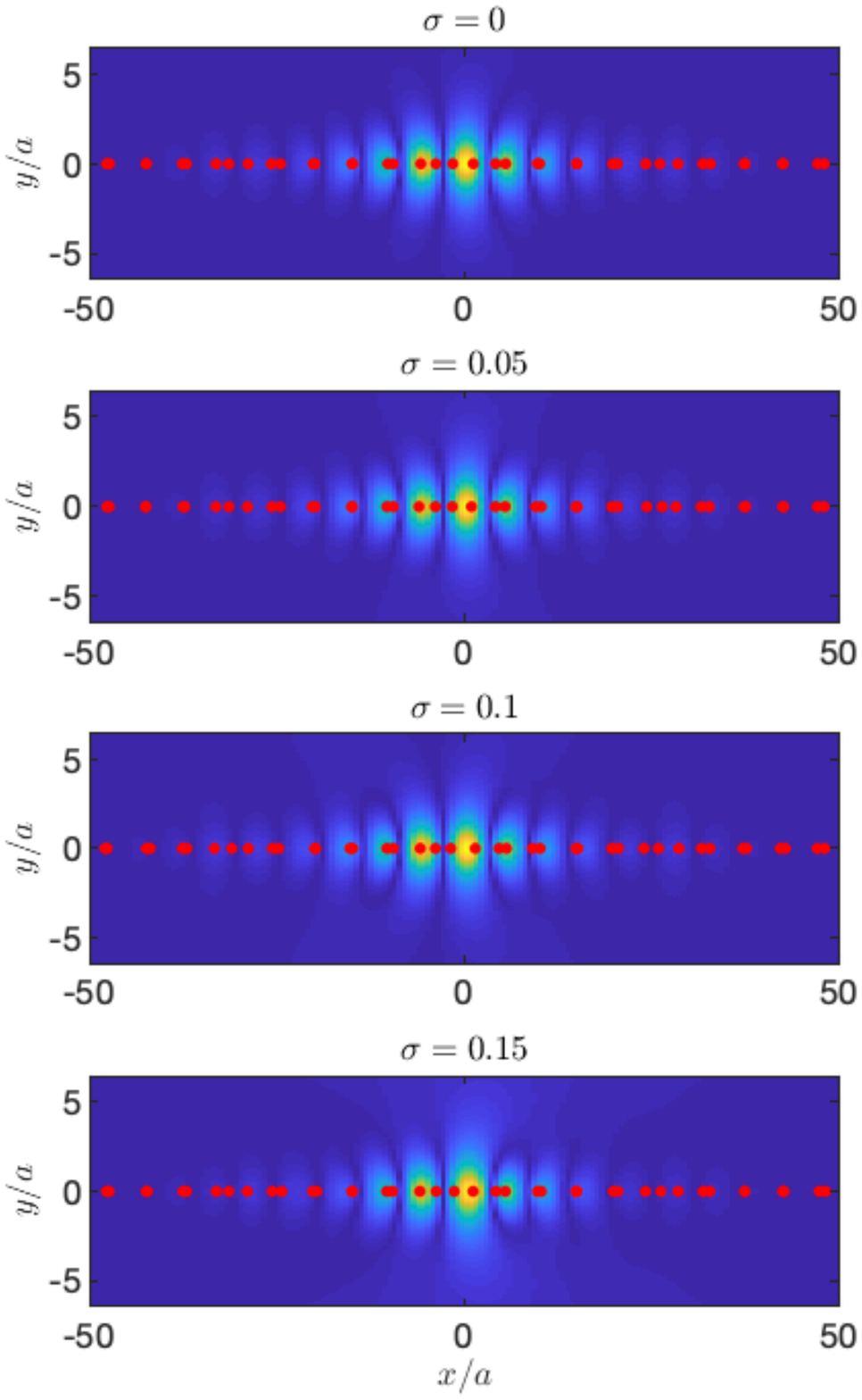}
	\caption{Spatial distribution of the localized mode found around $\Omega a = 0.19$ for some clusters with different disorders applied over them. All the modes have been normalized to the maximum field scattered by the non-disordered mode. The shape of the mode does not change with the disorder applied to the structure. When disorder magnitude approaches the amplitude of the modulation applied, the spatial distribution of the mode changes, and it is no longer an edge mode.}
	\label{Figure6}
\end{figure}

Figure \ref{Figure6} shows the edge state located around $\Omega a=0.19$ for some of the disordered configurations. The robustness and localization of this mode is clear from these plots, since the change in the shape of the field distribution is imperceptible. 

%%%%%%%%%%%%%%%%%%%%%%%%%%%%%%%%%%%%%%%%%%%%%%%%%%%%%%%%%%%%%%%%%%%%%%%%
\section{\label{sec:summary}Summary}
%%%%%%%%%%%%%%%%%%%%%%%%%%%%%%%%%%%%%%%%%%%%%%%%%%%%%%%%%%%%%%%%%%%%%%%%
In summary, we have shown that quasi-periodic lines of scatterers are capable of trapping flexural waves in two-dimensions. We have employed multiple scattering theory for the analysis of these structures, avoiding in this way the use of super-cell methods which artificially introduce periodicity in the clusters. Mapping the spectrum of these clusters as a function of the quasi-periodic modulation generates the Hofstadter's butterfly.

We have also shown that finite clusters built on a quasi-periodic pattern support interface states, enhanced when the clusters are merged with their chiral versions. Moreover, we have analyzed the bound states in clusters where a positional disorder has been introduced, and we have found that the modes are robust in the sense that their frequency remains unaltered as well as their spatial distribution.

The advantage of this geometry is that the bound state is not surrounded by the bulk material, since it is a zero-dimensional mode, induced by a one-dimensional material in a two-dimensional space, what makes this geometry more suitable for applications where propagating waves are expected to excite these modes, like those related with surface acoustic waves sensors. Since the methods developed in this work are general and not unique of flexural waves, we expect similar results for other mechanical waves as well as for electromagnetic waves.

\begin{acknowledgments}
Dani Torrent acknowledges financial support through the ``Ram\'on y Cajal'' fellowship under grant number RYC-2016-21188 and to the Ministry of Science, Innovation and Universities through Project No. RTI2018- 093921-A-C42. Marc Mart\'i-Sabat\'e acknowledges financial support through the FPU program under grant number FPU18/02725.
\end{acknowledgments}

\section*{Data Availability Statement}
The data that support the findings of this study are available from the corresponding author upon reasonable request.

%\bibliography{aipsamp}% Produces the bibliography via BibTeX.


\end{document}